\documentclass{article}
\usepackage{spconf,amsmath,graphicx,hyperref}
\usepackage{booktabs}
\usepackage{amssymb}
\usepackage{pifont}
\usepackage{multirow}
\usepackage{subcaption}
\usepackage{balance}

\title{VoxMorph: Scalable Zero-shot Voice Identity Morphing via Disentangled Embeddings}

\name{Bharath Krishnamurthy and Ajita Rattani}
\address{University of North Texas, Denton, Texas, USA \\
\small{\texttt{\{BharathKrishnamurthy@my.unt.edu, Ajita.Rattani@unt.edu\}}}}

\begin{document}
\ninept
\maketitle

\begin{abstract}
Morphing techniques generate artificial biometric samples that combine features from multiple individuals, allowing each contributor to be verified against a single enrolled template. While extensively studied in face recognition, this vulnerability remains largely unexplored in voice biometrics. Prior work on voice morphing is computationally expensive, non-scalable, and limited to acoustically similar identity pairs, constraining practical deployment. Moreover, existing sound-morphing methods target audio textures, music, or environmental sounds and are not transferable to voice identity manipulation. We propose VoxMorph, a zero-shot framework that produces high-fidelity voice morphs from as little as five seconds of audio per subject without model retraining. Our method disentangles vocal traits into prosody and timbre embeddings, allowing fine-grained interpolation of speaking style and identity. These embeddings are fused via Spherical Linear Interpolation (Slerp) and synthesized using an autoregressive language model (LM) coupled with a Conditional Flow Matching (CFM) network. VoxMorph achieves state-of-the-art performance, delivering a $2.6\times$ gain in audio quality, a $73\%$ reduction in intelligibility errors, and a $67.8\%$ morphing attack success rate on automated speaker verification (ASV) systems under strict security thresholds, establishing a practical and scalable paradigm for voice morphing with significant implications for biometric security. The code and dataset are available on our project page:~\href{https://vcbsl.github.io/VoxMorph/}{Vcbsl/VoxMorph}.
\end{abstract}

\begin{keywords}
Voice morphing, text-to-speech, zero-shot learning, speaker embedding, interpolation, speech synthesis
\end{keywords}


\begin{figure*}[t]
  \centering
  \includegraphics[width=0.85\linewidth]{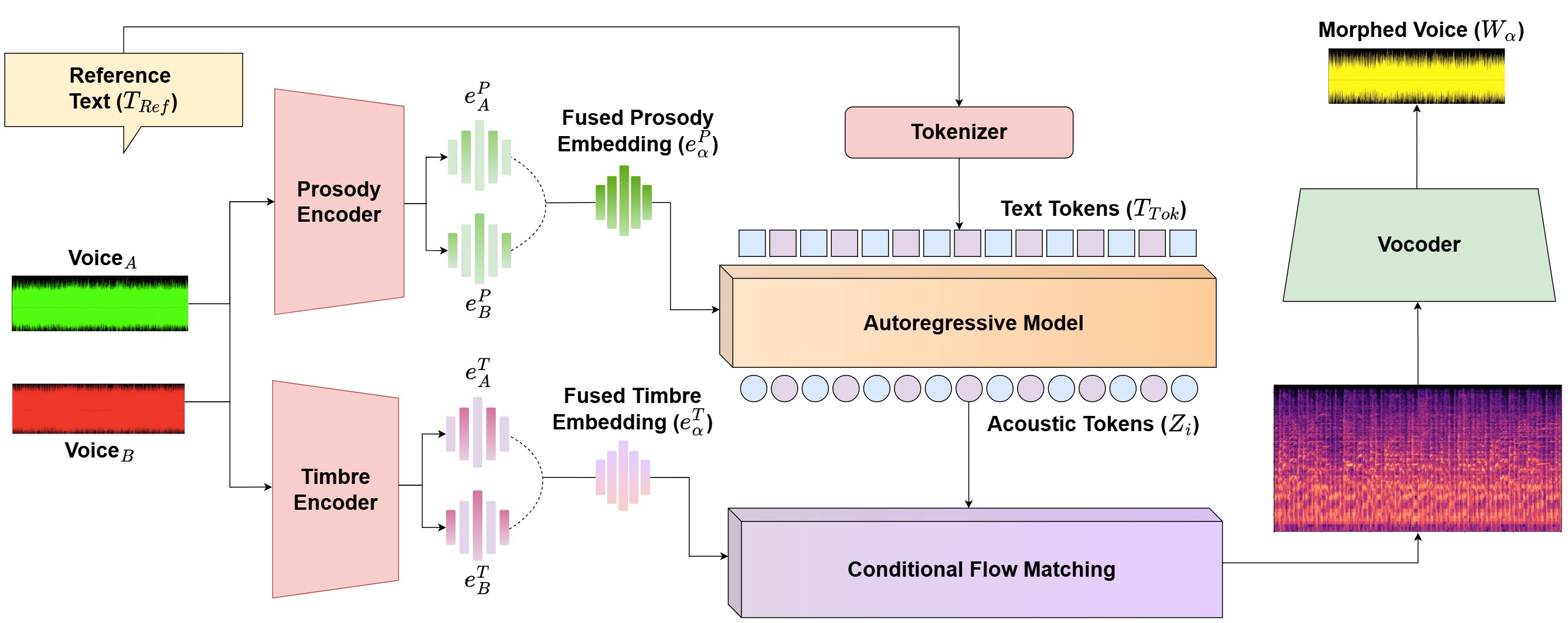}
  \caption{Architectural overview of the VoxMorph framework. The process consists of three core stages: (1) \textbf{Extraction:} Disentangled prosody (style) and timbre (identity) embeddings are extracted from two speaker identities. (2) \textbf{Interpolation:} The embedding representations are independently interpolated using Slerp. (3) \textbf{Synthesis/ Morphing:} The fused prosody embedding conditions an autoregressive language model, while the fused timbre embedding guides a flow matching network to generate a mel-spectrogram. A neural vocoder then converts the spectrogram into a high-fidelity morphed waveform.}
  \label{fig:audio_morph_workflow}
  \vspace{-4mm}
\end{figure*}

\section{Introduction}
\label{sec:intro}

The field of generative speech synthesis has achieved remarkable realism, with modern Text-to-Speech (TTS) models generating audio that is often indistinguishable from human speech~\cite{casanova2022yourtts, du2024cosyvoice, zhang2025vevo, chatterboxtts2025}. While this progress enables numerous creative applications, such as personalized voice assistants and expressive content creation~\cite{kim2025adaptvc}, they have also introduced significant risks to the security of Automatic Speaker Verification (ASV) systems~\cite{wu2015spoofing} due to deepfakes, a problem actively studied in challenges like ASVspoof~\cite{liu2023asvspoof,pujari2025waveverify}.


Apart from deepfakes, morphing attacks pose a significant threat to biometric security. Unlike Voice Conversion (VC), which maps a source voice to a single target identity for impersonation, Voice Identity Morphing (VIM) aims to create a synthetic biometric sample that can be successfully verified as belonging to two or more identities~\cite{Magic2014Ferrara}. This dual-identity claim makes VIM a uniquely insidious threat compared to standard spoofing. While morphing has been well studied for face and ocular biometrics~\cite{scherhag2017biometric, zhang2021mipgan, venkatesh2021face, krishnamurthy2025doomgan}, it remains largely unexplored in the voice domain. However, only a single study has addressed voice morphing~\cite{pani2023voice}, leaving a critical gap in our understanding of how such attacks affect the vulnerabilities of ASV systems. A distinction must be made between voice morphing~\cite{pani2023voice}, which involves the precise manipulation of human identity–specific biometric features, and general-purpose sound morphing, which has demonstrated success in transforming audio textures, environmental sounds, and music~\cite{gupta2023towards, kamath2025morphfader, niu2024soundmorpher}. 

The foundational work by Pani et al.~\cite{pani2023voice} is the only study demonstrating the feasibility of Voice Identity Morphing (VIM) attacks against ASV systems. However, their pioneering method suffers from significant practical and methodological limitations that prevent it from serving as a scalable threat. Specifically, the approach is not zero-shot, requiring an expensive, identity pair–specific fine-tuning process that demands over $30$ minutes of data per speaker and $8$–$10$ hours of training to generate a single morph. In addition, its reported success depends on careful pre-selection of acoustically similar target identity pairs, impacting its generalizability. These constraints, limited scalability and lack of zero-shot capability, render the approach impractical for producing biometrically coherent voice morphs at scale.
 
Another limitation of this pioneering approach lies in its use of a single, \textit{monolithic speaker embedding vector} to represent all of a speaker’s vocal characteristics, i.e., timbre and prosody, making it difficult to control each feature independently. Further, when embeddings from multiple speaker identities are combined at this level, the resulting voice morphs often exhibit subtle inconsistencies that modern Morphing Attack Detection (MAD) systems can easily detect, as these systems are designed to identify synthetic or manipulated speech.

We introduce \textbf{VoxMorph}, a novel framework that advances the state-of-the-art in voice identity morphing through a principled disentanglement of vocal representations into timbre and prosody. Our methodology leverages a key principle from state-of-the-art TTS architectures~\cite{chatterboxtts2025, du2024cosyvoice2} based on the sequential separation of speech generation into distinct stages. This allows us to enable the decomposition of a voice into its core components: a \textit{prosody embedding} that encapsulates the speaking style~\cite{wan2018generalized, jia2018transfer} and a \textit{timbre embedding} that encodes the core vocal identity~\cite{wang2023cam++}. This disentanglement is advantageous because it allows for the independent manipulation of vocal style and identity information, preventing the acoustic artifacts common in the prior VIM model, enabling more granular, high-fidelity voice morphs. Further, instead of simple linear averaging, we employ a Slerp mechanism to independently fuse disentangled embeddings across multiple identities. The resulting fused embeddings then condition a three-stage synthesis pipeline to produce the final morphed audio waveform. This disentangled approach facilitates the generation of perceptually seamless voice morphs and overcomes the acoustic artifacts inherent in the VIM model. The summarized \textbf{contributions} of our work are as follows:


\begin{enumerate}
    \item We introduce VoxMorph, a zero-shot framework that overcomes the limitations of the prior VIM model, dependent on large amounts of identity-specific data and costly fine-tuning for each new speaker identity pair.
    \item We demonstrate the principled use of Slerp to independently fuse prosody and timbre embeddings across identities, which in turn guide an autoregressive LM and a CFM network to synthesize the final waveform.  This disentangled approach mitigates the acoustic artifacts common to the prior VIM model.
    \item We present a highly efficient and practical method that constitutes a scalable attack vector, requiring only 5 seconds of source audio and no model retraining for each new pair of speakers, a fundamental advance over the existing voice morphing technique.
    \item We establish a new \textbf{state-of-the-art} in voice morphing, achieving superior audio quality and, most critically, an unprecedented 67.8\% morphing attack success rate on an ASV system operating at stringent thresholds. Further, we release the first publicly available dataset of $10,000$ high-fidelity voice morphs to develop next-generation countermeasures.
\end{enumerate}

\section{Background}
\label{sec:background}

The vulnerability of morphing is well-documented and extensively explored in visual domains like face, iris, and ocular biometrics~\cite{zhang2021mipgan, sharma2021image, krishnamurthy2025doomgan}. In stark contrast, the audio domain remains critically underexplored. While techniques exist for general-purpose sound morphing, they target creative applications such as transforming audio textures~\cite{gupta2023towards} or environmental sounds~\cite{kamath2025morphfader, niu2024soundmorpher}. These methods are fundamentally unsuited for biometric attacks because these methods focus on acoustic qualities rather than speaker identity. On the other hand, zero-shot voice imitation (conversion) systems like Vevo~\cite{zhang2025vevo} aim to perform voice conversion from the source speaker to the target voice. However, these powerful models are designed for transformations from a single source identity to a single target identity and do not inherently support the blending of two or more source identities into a single morphed output. 

To date, the foundational study in VIM by Pani et al.~\cite{pani2023voice} demonstrated the possibility of such an attack on ASV systems. However, their pioneering method is critically limited by its computational expensiveness, as the method requires large amounts of data (over 30 minutes per speaker) and time (8-10 hours per morph), making it impractical as a scalable threat. This research gap remains despite the availability of foundational tools enabled by advances in generative speech synthesis. Modern zero-shot TTS models, exemplified by YourTTS~\cite{casanova2022yourtts, du2024cosyvoice, du2024cosyvoice2, chatterboxtts2025}, can clone a speaker’s voice from just a few seconds of audio by conditioning a generative network on a speaker embedding. Building on this paradigm, we propose a novel method to generate synthetic voice morphs that leverages the capabilities of these modern TTS systems.

\begin{table*}[t]
\vspace{-4mm}
\centering
\small
\renewcommand{\arraystretch}{0.9}
\setlength{\tabcolsep}{7pt}
\begin{tabular}{l|cccc|ccc|ccc}
\hline
\multirow{2}{*}{\textbf{Method}} & \textbf{FAD \textdownarrow} & \textbf{FAD \textdownarrow} & \textbf{KLD} & \textbf{WER} & \multicolumn{3}{c|}{\textbf{MMPMR (\%)}} & \multicolumn{3}{c}{\textbf{FMMPMR (\%)}} \\
\cline{6-11} 
& (Vs Real) & (Vs Clone) & (\textdownarrow) & (\textdownarrow) & \textbf{@ 0.01\%} & \textbf{@ 0.1\%} & \textbf{@ 1\%} & \textbf{@ 0.01\%} & \textbf{@ 0.1\%} & \textbf{@ 1\%} \\
\hline
MorphFader \cite{kamath2025morphfader} & 8.96 & 0.25 & 0.4332 & 1.84 & 0.0 & 0.0 & 0.0 & 0.0 & 0.0 & 0.0 \\ 
Vevo \cite{zhang2025vevo} & 9.14 & 0.63 & 0.1899 & 0.54 & 82.40 & 94.60 & 98.80 & 9.00 & 44.00 & 85.60 \\ 
ViM \cite{pani2023voice} & 7.52 & 1.52 & 0.3501 & 1.06 & 2.61 & 29.66 & 89.38 & 0.00 & 5.61 & 52.10 \\ 
\textbf{VoxMorph-v1} & 5.03 & \textbf{0.24} & 0.1404 & 0.33 & 78.60 & 98.40 & 100 & 60.60 & 96.00 & 99.80 \\ 
\textbf{VoxMorph-v2} & \textbf{4.90} & 0.27 & \textbf{0.1385} & \textbf{0.19} & \textbf{99.80} & \textbf{100} & \textbf{100} & \textbf{67.80} & \textbf{97.20} & \textbf{100} \\
\hline
\end{tabular}
\caption{Quantitative comparison with state-of-the-art audio morphing methods. Our method demonstrates superior performance in quality (FAD, KLD), intelligibility (WER), and morphing attack success (MMPMR/FMMPMR). Note that for FAD, KLD, and WER, lower is better.}
\label{tab:main_results}
\vspace{-4mm}
\end{table*}

\section{Proposed Framework}
\label{sec:method}
We introduce a novel, zero-shot voice morphing methodology designed to create high-fidelity vocal interpolations, with the complete pipeline illustrated in Figure~\ref{fig:audio_morph_workflow}. Our approach builds upon the multi-stage design of modern cascaded Text-to-Speech (TTS) architectures~\cite{chatterboxtts2025, du2024cosyvoice2} by first decomposing the voices of two source speakers into their fundamental components. Specifically, we employ two specialized encoders to extract disentangled representations: a \textit{prosody embedding} for speaking style and a \textit{timbre embedding} for the core vocal identity. The proposed framework then uses Slerp to blend the prosody and timbre representations separately. The synthesis stage is then conditioned on these fused embeddings: the fused prosody embedding guides an autoregressive LM, while the fused timbre embedding conditions a flow matching network to produce a mel-spectrogram. Finally, a neural vocoder synthesizes the high-fidelity waveform. This disentangled approach affords granular and perceptually smooth control over the vocal transition without requiring any speaker-specific fine-tuning. This process is discussed in detail below:

\vspace{-2mm}

\subsection{Disentangled Vocal Feature Extraction}


The foundation of our framework is the disentanglement of a speaker's voice into prosody and timbre from minimal duration audio sample ($\geq$5s). We employ a dual-encoder architecture where the first encoder, based on the GE2E architecture~\cite{wan2018generalized}, captures the high-level speaking \textit{style} (rhythm, pitch) into a \textbf{prosody embedding} ($\mathbf{e}^{P}_i$). Concurrently, to ensure robust identity representation, we utilize a CAM++ encoder~\cite{wang2023cam++} to extract the core biometric \textit{identity} into a \textbf{timbre embedding} ($\mathbf{e}^{T}_i$), where $i \in \{A, B\}$ denotes one of the two source speaker identities.


\subsection{Interpolation}
The prosody and timbre embeddings of the two source speakers, $\mathbf{e}^{P}_A, \mathbf{e}^{P}_B$ and $\mathbf{e}^{T}_A, \mathbf{e}^{T}_B$, are independently fused using Slerp interpolation to obtain the fused embeddings, $\mathbf{e}^{P}_\alpha$ and $\mathbf{e}^{T}_\alpha$, controlled by an interpolation factor $\alpha \in [0,1]$:
\begin{equation}
\mathbf{e}^{(X)}_\alpha = \frac{\sin((1-\alpha)\Omega)}{\sin(\Omega)}\mathbf{e}^{(X)}_A + \frac{\sin(\alpha\Omega)}{\sin(\Omega)}\mathbf{e}^{(X)}_B,
\end{equation}
where $\Omega = \arccos(\mathbf{e}^{(X)}_A \cdot \mathbf{e}^{(X)}_B)$ is the angle between the embeddings of a given type ($X \in \{P,T\}$). This independent interpolation preserves the geometric structure of the hypersphere, minimizing audio artifacts while allowing granular control over style and identity. For all experiments reported in this work, we set the interpolation factor $\alpha \geq 0.5$ to ensure an equal blend of both source identities. 

\subsection{Multi-Stage Synthesis of the Morphed Voice}

The fused prosody and timbre embeddings, $\mathbf{e}^{P}_\alpha$ and $\mathbf{e}^{T}_\alpha$, are then passed through a three-stage synthesis pipeline to generate the final morphed waveform, $W_\alpha$. Prosody guides the stylistic structure in the acoustic token generation stage, while timbre controls the vocal identity in the mel-spectrogram synthesis stage, producing the final morphed vocal identity as follows:

\begin{enumerate}
    \item \textbf{Acoustic Token Generation:} An autoregressive LM~\cite{touvron2023llama} generates a sequence of discrete acoustic tokens, $\mathbf{z}$, from the input text. The \textit{fused prosody embedding}, $\mathbf{e}^{P}_\alpha$, is projected into the token space and prepended as a conditioning prefix to the input text sequence, guiding the model to synthesize the acoustic structure in the target morph style.
    \begin{equation}
        z_i \sim P(z_i | \mathbf{z}_{<i}, \mathbf{T}_{tok}, \mathbf{e}^{P}_\alpha)
    \end{equation}

    \item \textbf{Mel-Spectrogram Synthesis:} A Conditional Flow Matching (CFM) model~\cite{du2024cosyvoice2} synthesizes the final mel-spectrogram ($\mathbf{Mel}$) from the acoustic tokens. This stage renders the core vocal identity by conditioning on the \textit{fused timbre embedding}, $\mathbf{e}^{T}_\alpha$. The model learns a vector field to solve a probability flow ODE, transforming a noise prior into the target spectrogram:
    \begin{equation}
        \mathbf{Mel} = \mathbf{x}_0 + \int_{0}^{1} \tilde{\mathbf{v}}_t \,dt
    \end{equation}
    where $\tilde{\mathbf{v}}_t$ is the guided vector field. Both of these stages employ Classifier-Free Guidance (CFG)~\cite{ho2022classifier} to improve adherence to the conditioning embeddings.

    \item \textbf{Waveform Synthesis:} Finally, a HiFTNet vocoder~\cite{li2023hiftnet} takes the generated mel-spectrogram ($\mathbf{Mel}$) and synthesizes the high-fidelity output morphed waveform, $W_\alpha$.
\end{enumerate}

\section{Experimental Setup}
\label{sec:setup}
\vspace{-1mm}

\subsection{Datasets and Implementation}
\label{sec:datasets}
\vspace{-1mm}
We validate the efficacy of our model on the publicly accessible librispeech dataset~\cite{panayotov2015librispeech} containing large-scale corpus of approximately 1000 hours of 16kHz read English speech. Specifically, we use the clean subset with over 100 hours of training data as the source speech. To create realistic voice morphs, we select female pairs and male pairs based on the available speaker information. Unlike previous research~\cite{pani2023voice} that use similarity scores computed with speaker verification systems~\cite{desplanques2020ecapa} to select the top 100 pairs for voice morphing, we demonstrate the efficacy of our method by skipping this step to select 500 random voice pairs for similar genders. For additional experimental results, readers are recommended to visit the project page.

\vspace{-1mm}
\subsection{Baselines}
\vspace{-1mm}
To rigorously evaluate our framework, we benchmark against three latest state-of-the-art voice and sound morphing systems, each representing a distinct domain of audio synthesis, replicated from their official codebases.  Our primary comparison is against \textbf{ViM}\cite{pani2023voice}, the foundational method for voice identity morphing. For this baseline, we replace the original DeepTalk encoder with the more widely adopted ECAPA-TDNN model\cite{desplanques2020ecapa}, an alternative suggested by the original authors due to its public availability. For the sake of completeness, we also benchmark \textbf{Vevo}\cite{zhang2025vevo}, a state-of-the-art zero-shot voice imitation model. Finally, to situate our work in the broader context of audio manipulation, we compare against \textbf{MorphFader}\cite{kamath2025morphfader}, a leading general-purpose sound morphing technique. 


\vspace{-1mm}
\subsection{Evaluation Metrics and Implementation Details}
\vspace{-1mm}
Performance is assessed using a comprehensive suite of metrics targeting audio quality, intelligibility, and morphing success. Audio quality is measured using the \textbf{Fréchet Audio Distance (FAD)}~\cite{kilgour2018fr} and \textbf{Kullback-Leibler Divergence (KLD)}. We compute FAD in two ways: against a diverse corpus of real speech to measure overall naturalness, and against high-quality voice clones of the source speakers to specifically assess the acoustic fidelity of the morphing process itself. Both FAD variants use VGGish embeddings. KLD measures spectral fidelity by comparing the normalized histograms of log Mel-spectrograms between a generated sample and its real source audio. Intelligibility is quantified by the \textbf{Word Error Rate (WER)}, computed using a pre-trained Wav2Vec2-Base-960h ASR model~\cite{baevski2020wav2vec}. The critical success of identity morphing is evaluated using two specialized metrics computed with an open-source Resemblyzer ASV system~\cite{Resemblyzer}. The \textbf{Mated Morphed Presentation Match Rate (MMPMR)} measures the proportion of morphs that successfully verify against \textit{at least one} of the source identities, while the stricter \textbf{Fully Mated Morphed Presentation Match Rate (FMMPMR)} requires successful verification against \textit{both} sources~\cite{ferrara2022morphing}. For our method, we evaluate two variants: \textbf{VoxMorph-v1}, using a single 5–20s clip per speaker, and \textbf{VoxMorph-v2}, which leverages up to seven clips ($\approx$ 1–2 min) for a more robust voice profile. All experiments were conducted on a single NVIDIA RTX 5000 Ada GPU.

\vspace{-2mm}

\section{Results and Analysis}
\label{sec:results}
\vspace{-3mm}

\subsection{Quantitative Comparison with Baselines}
\label{ssec:baseline}
\vspace{-2mm}
Table~\ref{tab:main_results} presents a comprehensive quantitative evaluation of our proposed method against state-of-the-art baselines across multiple dimensions of voice morphing quality. VoxMorph establishes new performance benchmarks across all evaluation metrics, demonstrating substantial improvements in audio quality, intelligibility, and morphing effectiveness.

\textbf{Audio Quality and Fidelity:} Our method achieves exceptional perceptual quality with FAD scores of 0.24 and 0.27 for single-clip and multi-clip variants, respectively. This represents a dramatic \textbf{6.3$\times$ improvement} over ViM~\cite{pani2023voice} (1.52) and a \textbf{2.6$\times$ improvement} over the previous best-performing Vevo baseline~\cite{zhang2025vevo} (0.63). Similarly, our KLD scores (0.1404 and 0.1385) demonstrate \textbf{2.5$\times$ better} spectral fidelity than ViM (0.3501) and \textbf{35\% improvement} over Vevo (0.1899), indicating that our morphed voices exhibit superior spectral consistency and are perceptually closer to natural human speech.

\textbf{Speech Intelligibility:} Our approach achieves remarkable intelligibility with WER scores of 0.33 with VoxMorph-v1 and 0.19 with VoxMorph-v2, representing a \textbf{73\% improvement} over the best baseline Vevo (0.54) and a \textbf{5.6\%improvement} over ViM (1.06). The near-perfect intelligibility of our multi-clip variant (WER: 0.19) demonstrates that our morphing process preserves linguistic content without the acoustic degradation common in existing methods.

\textbf{Morphing Effectiveness:} The most compelling evidence of our method's superiority lies in the morphing attack success metrics. At the strictest evaluation threshold (0.01\% FAR), our single-clip variant achieves 78.6\% MMPMR, dramatically outperforming ViM's \textbf{30$\times$ lower} performance (2.61\%). More critically, our FMMPMR scores, measuring successful morphing, reveal the fundamental limitations of existing approaches. While ViM completely fails at strict thresholds (0.00\% FMMPMR at 0.01\%), our method achieves 60.6\% with VoxMorph-v1, while VoxMorph-v2 achieves 67.8\%. Vevo, despite reasonable MMPMR scores, achieves only 9.00\% FMMPMR at the strictest threshold, while MorphFader, a model designed for sound morphing, fails to translate its success to voice identity morphing.


\textbf{Scalability and Practical Advantages:} Unlike ViM~\cite{pani2023voice}, which requires over 30 minutes of data per speaker and 8–10 hours of fine-tuning per morph, our zero-shot method generates high-quality morphs from only 5–20 seconds of audio without retraining. This shift from pair-specific training to a scalable, plug-and-play framework is enabled by disentangling prosody and timbre into separate embeddings to generate high-fidelity voice morphs.

\begin{table}[t]
\vspace{-4mm}
\centering
\small
\renewcommand{\arraystretch}{0.9}
\setlength{\tabcolsep}{3pt}
\begin{tabular}{lcccccc}
\toprule
\multirow{2}{*}{\textbf{Interpolation Method}} & \textbf{FAD\textdownarrow} & \textbf{WER} & \textbf{KLD} & \multicolumn{2}{c}{\textbf{FMMPMR(\%)}} \\
\cmidrule(lr){5-6}
& \textbf{(vs Real)} & \textbf{\textdownarrow} & \textbf{\textdownarrow} & \textbf{@ 0.01} & \textbf{@ 0.1} \\
\midrule
Linear Averaging      & 5.03 & \textbf{0.1917} & 0.1854 & 34.80 & 83.80 \\
Lerp                  & 4.96 & 0.1928 & 0.1838 & 62.60 & 96.80 \\
Lerp (P) + Slerp (T) & 5.58 & 0.1971 & 0.1826 & 62.80 & 96.00 \\
Slerp (P) + Lerp (T) & 5.06 & 0.1926 & 0.1813 & 62.60 & \textbf{97.20} \\
\midrule
\textbf{Slerp (VoxMorph)} & \textbf{4.90} & 0.1920 & \textbf{0.1385} & \textbf{67.80} & \textbf{97.20} \\
\bottomrule
\end{tabular}
\vspace{-3mm}
\caption{Ablation study of interpolation methods. The Slerp approach shows superior performance with lower FAD, WER, and KLD values, and a substantially higher FMMPMR, indicating a more successful morphing of speaker identities.}
\label{tab:ablation_results}
\vspace{-2mm}
\end{table}

\begin{table}[t]
\vspace{-2mm}
\centering
\small
\renewcommand{\arraystretch}{0.8}
\setlength{\tabcolsep}{7pt}
\begin{tabular}{lcccc}
\toprule
\multirow{2}{*}{\textbf{Method}} & \multicolumn{2}{c}{\textbf{MMPMR (\%)}} & \multicolumn{2}{c}{\textbf{FMMPMR (\%)}} \\
\cmidrule(lr){2-3} \cmidrule(lr){4-5}
& \textbf{@ 0.01\%} & \textbf{@ 0.1\%} & \textbf{@ 0.01\%} & \textbf{@ 0.1\%} \\
\midrule
GE2E     & \textbf{78.60}     & 98.40     & \textbf{60.60}      & \textbf{96.00} \\
ECAPA    & 72.20     & 98.40     & 48.20      & 93.00 \\
HUBERT   & 75.40     & \textbf{99.00}     & 48.40      & 91.20 \\
Wav2Vec2 & 75.80     & 98.80     & 47.60      & 90.60 \\
\bottomrule
\end{tabular}
\vspace{-3mm}
\caption{Ablation study of different prosody encoder models. The GE2E encoder demonstrates a significantly higher MMPMR and FMMPMR at the strict 0.01\% threshold, indicating its superior capability for both impersonation and identity morphing.}
\label{tab:ablation_results_encoder}
\vspace{-4mm}
\end{table}

\vspace{-2mm}
\subsection{Ablation Study}
\label{ssec:ablation}
\vspace{-2mm}
We conducted two ablation studies to validate our core design choices. First, we evaluated the Slerp strategy against linear averaging, Linear Interpolation (Lerp), and mixed Lerp/Slerp approaches. As shown in Table~\ref{tab:ablation_results}, while all methods produced intelligible speech, Slerp demonstrated clear superiority in identity morphing. At the strict 0.01\% FAR threshold, it achieved an FMMPMR of 67.80\%, a significant 5.2\% absolute improvement over the Lerp strategy. This advantage stems from the geometric properties of the L2-normalized embeddings, which reside on a hypersphere. Slerp computes the geodesic path along this manifold, ensuring the interpolated vector remains a valid, high-quality speaker representation, thereby preventing the spectral degradation and identity loss associated with off-manifold linear methods. Secondly, we analyzed the impact of different prosody encoder architectures by substituting the GE2E~\cite{wan2018generalized} model with ECAPA-TDNN~\cite{desplanques2020ecapa}, HuBERT~\cite{hsu2021hubert}, and Wav2Vec2~\cite{baevski2020wav2vec}. The results in Table~\ref{tab:ablation_results_encoder} show that while all encoders perform well, the LSTM-based GE2E architecture is the most effective for creating a convincing identity morphing. It achieved a superior FMMPMR of 60.60\% at the 0.01\% threshold, outperforming the next-best models by over 12\%. We hypothesize that the recurrent nature of GE2E is inherently better suited to capturing the dynamic, time-varying characteristics of prosody (e.g., rhythm, pitch). This leads to a more structured latent space that is more amenable to Slerp's geodesic interpolation, resulting in a more biometrically coherent morphing of the source identities.

\vspace{-3.5mm}
\section{Conclusion}
\label{sec:conclusion}
\vspace{-3.5mm}
We have presented a novel framework for zero-shot voice identity morphing that achieves state-of-the-art performance by operating on disentangled vocal embeddings. By disentangling voice characteristics into prosody and timbre and interpolating them independently using a Slerp mechanism, our method produces seamless, high-fidelity vocal transitions from minimal-duration voice data without any fine-tuning. Our experimental results confirm significant improvements in audio quality, intelligibility, and morphing effectiveness over existing research. This work establishes voice morphing as a potent and scalable threat to ASV systems. By presenting a framework that significantly advances the state-of-the-art in voice morph generation, we provide the research community with a critical tool to develop and benchmark next-generation MAD countermeasures. The release of our large-scale dataset is intended to accelerate this effort. Future research will extend this framework in several key directions. A primary focus will be on investigating the morphing of more than two identities into a single, coherent voiceprint. Additionally, we will explore the application of this technique to cross-lingual voice morphing and its potential for real-time synthesis. 


\balance
\bibliographystyle{IEEEbib}
\bibliography{refs_final}

\end{document}